\documentclass{svjour3}                     
\smartqed  
\usepackage{graphicx}
\usepackage{epsfig}
\usepackage{amsmath}
\usepackage{amsfonts}

\newcommand{\bfsigma}{\mbox{\boldmath $\sigma$}}
\newcommand{\bflambda}{\mbox{\boldmath $\lambda$}}
\newcommand{\bfrho}{\mbox{\boldmath $\rho$}}

\newcommand{\bx}{\boldsymbol{x}}
\newcommand{\br}{\boldsymbol{r}}
\newcommand{\brg}{\boldsymbol{R}}
\newcommand{\lQ}{\Lambda_{\rm QCD}}

\newcommand{\als}{\alpha_{\rm s}}
\newcommand{\MS}{\overline{\rm MS}}
\newcommand{\siml}{{\ \lower-1.2pt\vbox{\hbox{\rlap{$<$}\lower6pt\vbox{\hbox{$\sim$}}}}\ }} 
\newcommand{\simg}{{\ \lower-1.2pt\vbox{\hbox{\rlap{$>$}\lower6pt\vbox{\hbox{$\sim$}}}}\ }}

\journalname{Few-Body Systems}

\begin{document}

\title{Effective field theories for baryons with two- and three-heavy quarks\thanks{
``Relativistic Description of Two- and Three-Body Systems in Nuclear Physics'', 
ECT*, October 19-23 2009}
}


\author{Antonio Vairo}


\institute{A. Vairo \at
Physik-Department, Technische Universit\"at M\"unchen, 
James-Franck-Str. 1, 85748 Garching, Germany \\
              \email{antonio.vairo@tum.de}           
}

\date{Received: date / Accepted: date}

\maketitle

\begin{abstract}
Baryons made of two or three heavy quarks can be described 
in the modern language of non-relativistic effective field theories. 
These, besides allowing a rigorous treatment of the systems, 
provide new insight in the nature of the three-body interaction in QCD.
\end{abstract}

\section{Motivations}
\label{sec_moti}
Baryons made of two or three heavy quarks offer an interesting alternative 
to quarkonium for studying the dynamics of non-relativistic systems in QCD 
and for investigating the transition region from Coulombic to confined bound states.

The modern approach to quarkonium physics consists in taking  
advantage of the hierarchy of non-relativistic energy scales in the system 
by constructing a suitable hierarchy of effective field theories (EFTs)~\cite{Brambilla:2004jw}.
The energy scales are the heavy-quark mass, $m$, the typical momentum transfer, 
$p \ll m$, whose inverse sets the typical distance, $r$, between the heavy quark and the antiquark, 
and the typical kinetic energy, $E \ll p$, whose inverse sets the typical time scale of the 
bound state. In the ultimate EFT, obtained after integrating out gluons 
of energy and momentum of the order of $m$ and $p$, the interaction between heavy quarks 
is organized as an expansion in powers of $1/m$ and $r$. 
At zeroth-order in $r$, the interaction is entirely encoded in the quark-antiquark potential, 
which, at zeroth-order in $1/m$, reduces to the static potential.
Terms proportional to powers of $1/m$ and $r$ can be systematically added. 

An analogous hierarchy of energy scales characterizes also baryons that contain at least 
two heavy quarks. Hence, we may describe these systems by means of 
EFTs analogous to the ones suited for heavy quarkonium~\cite{Brambilla:2005yk,Fleming:2005pd}.
In particular, the ultimate EFT, obtained after integrating out gluons 
of energy and momentum of the order of the heavy-quark masses and of the typical 
momentum transfer between heavy quarks, is organized as an expansion 
in the inverse of the heavy-quark masses and in the distance between the heavy quarks. 
For equal heavy-quark masses, the structure of the Lagrangian and its 
power counting are similar to the quarkonium case; the Lagrangian reads 
\begin{equation}
\mathcal{L} = -\frac{1}{4} F^a_{\mu\nu}F^{a\mu\nu} 
+ \sum_f\bar{q}_f \, iD \!\!\!\!/\, q_f 
+ \sum_{i,j}\delta\mathcal{L}^{(i,j)},
\label{lagrangian}
\end{equation}
where $\delta\mathcal{L}^{(i,j)}$ are terms containing the heavy quark or antiquark fields, 
which are proportional to $1/m^i \times [\hbox{typical distance between heavy quarks/antiquarks}]^j$.
The fields $q_f$ are $n_f$ light-quark fields, assumed to be massless. 
The heavy quark or antiquark fields in $\delta\mathcal{L}^{(i,j)}$ are twice those in the bound state, 
e.g, in the quarkonium case, $\delta\mathcal{L}^{(i,j)}$ contains two quark and two antiquark fields.
For different heavy-quark masses, more scales are involved; 
in the following, we will not consider such cases.

Although the EFTs for quarkonium and baryons made of two or three heavy quarks 
are similar in structure, they are characterized by different degrees of freedom.
This is best seen when the typical distance between heavy quarks is smaller 
than the inverse of the  hadronic scale $\lQ$, which is the case that we will discuss in this note.
At these distances, gluons may resolve coloured degrees of freedom. These are gluons or light quarks 
or, in the quarkonium case, $Q\bar{Q}$ states in a colour singlet or in a colour octet configuration.
In the case of baryons made of two heavy quarks $Q$ and a light quark $q$, at distances 
smaller than $1/\lQ$, gluons can resolve gluons, light quarks and $QQ$ pairs 
in a colour antitriplet or in a colour sextet configuration.
It is the binding of the antitriplet with the light quark $q$ that forms the $QQq$ baryon.
The system very much resembles a heavy-light meson, with the heavy antiquark replaced by 
a $QQ$ antitriplet; this fact may be exploited to deduce
some properties of the $QQq$ baryons from the corresponding $\bar{Q}q$ mesons~\cite{Savage:di}.
In the case of baryons made of three heavy quarks $Q$, at distances  
smaller than $1/\lQ$, gluons can resolve gluons, light quarks and $QQQ$ states  
either in a colour singlet or in two different colour octets 
or in a colour decuplet configuration.

The lattice evaluation of the $QQQ$ static potential has a long tradition 
(see e.g.~\cite{Takahashi:2000te} and references therein), while 
the static potential between a $QQ$ pair in the presence of a light quark has been 
evaluated on the lattice only recently~\cite{Yamamoto:2007pf}. 
Expressions for the $1/m$ and the $1/m^2$ spin-dependent $QQQ$ 
potentials in terms of Wilson loops can be found in~\cite{Brambilla:2005yk}, 
but have not been calculated on the lattice yet (while the complete expressions 
of all the $Q\bar{Q}$ potentials up to order $1/m^2$ can be found in~\cite{Brambilla:2000gk} 
and the most recent lattice determinations are in~\cite{Koma:2006si}).
Perturbative studies of the static potential may help to understand the transition region 
from the perturbative to the non-perturbative regime. In the case of the 
$Q\bar{Q}$ potential, this region is characterized by the smooth transition 
from a Coulomb potential to a linear raising one (for recent studies, see~\cite{Brambilla:2010pp}). 
In the case of the $QQQ$ static potential, the transition from the perturbative 
to the non-perturbative regime is accompanied by the emergence of a three-body potential 
that depends on one length only (see, for instance, \cite{Takahashi:2000te}). 
This is a rather spectacular phenomenon, which has been investigated 
only recently from a perturbative perspective~\cite{Brambilla:2009cd}.

There is so far no experimental evidence of $QQQ$ baryons, while a few years ago the 
SELEX experiment has claimed evidence of possible doubly charmed baryon states~\cite{Mattson:2002vu}.
Until today this evidence has not been confirmed by other experiments, 
but is mostly behind the revival of interest in this kind of systems at the mid of this decade.

\section{EFT for $QQq$}
\label{sec_QQq}
The EFT Lagrangian that describes $QQq$ baryons below the momentum transfer scale $p$, assumed 
to be larger than $\lQ$, has the general form of Eq. \eqref{lagrangian}, with 
$\delta\mathcal{L}^{(i,j)}$ made out of four quark fields~\cite{Brambilla:2005yk}. 

The term $\delta\mathcal{L}^{(0,0)}$ is 
\begin{equation}
\delta\mathcal{L}^{(0,0)} = 
\int d^3r \; T^\dagger \left[ iD_0 - V_T^{(0)} \right] T 
+ \Sigma^\dagger \left[ iD_0 - V^{(0)}_\Sigma \right] \Sigma,
\end{equation}
where $T=(T^1,T^2,T^3)$ are the three independent $QQ$ antitriplet fields, 
$\Sigma = (\Sigma^1, \Sigma^2,$ $\dots, \Sigma^6)$ are the six independent $QQ$ sextet fields
and the gauge fields in the covariant derivatives acting on the 
antitriplet and sextet fields are understood in the $\bar{3}$ and $6$ 
representations respectively.
The matching coefficients $V_T^{(0)}$ and $V_\Sigma^{(0)}$ can be identified 
with the antitriplet and sextet static potentials respectively.
The $QQ$ antitriplet static potential $V_T^{(0)}$ has been calculated recently up to 
next-to-next-to-leading order (NNLO)~\cite{Brambilla:2009cd}; it reads
\begin{eqnarray}
V_T^{(0)} \!\! &=& \!\! -\frac{2}{3}
\frac{\als(1/\vert\br\vert)}{\vert\br\vert}
\left\{1+\frac{\als(1/\vert\br\vert)}{4\pi}\left[
\frac{31}{3}+22 \gamma_E -\left(\frac{10}{9}+\frac{4}{3}\gamma_E\right)n_f\right]
\right.
\nonumber\\
&& 
\hspace{5mm}
\left.
+\left(\frac{\als}{4\pi}\right)^2
\left[66 \zeta(3)+484 \gamma_E^2+ \frac{1976}{3}\gamma_E + \frac{3}{4}\pi^4 + \frac{121}{3}\pi^2 + \frac{4343}{18}
\right.\right.
\nonumber\\
&& \hspace{25mm}
-\left(\frac{52}{3} \zeta(3) + \frac{176}{3}\gamma_E^2 +\frac{916}{9}\gamma_E 
+ \frac{44}{9}\pi^2 + \frac{1229}{27}\right)n_f 
\nonumber\\
&& \hspace{25mm}
\left.\left.
+\left(\frac{16}{9}\gamma_E^2 + \frac{80}{27}\gamma_E + \frac{4}{27}\pi^2
+\frac{100}{81}\right)n_f^2
\right]
\right\},
\end{eqnarray}
where $\als$ is the strong-coupling constant in the $\MS$ scheme. 
The $QQ$ sextet static potential $V_\Sigma^{(0)}$ is repulsive at leading order:
$V_\Sigma^{(0)} = \als/(3\vert\br\vert)$.

Several terms contribute to $\delta\mathcal{L}^{(1,0)}$. The one responsible for the 
hyperfine splitting~is 
\begin{equation}
\delta\mathcal{L}^{(1,0)}_{\rm hfs} = 
\frac{V_{T\bfsigma\cdot{\boldsymbol{B}}T}^{(1,0)}}{2}
T^\dagger 
\frac{c_F}{2m} \left( \bfsigma^{(1)}  + \bfsigma^{(2)} \right) 
\cdot g{\boldsymbol{B}}^a  T^a_{\bar{3}}\, T,
\label{QQhfs}
\end{equation} 
where $V_{T\bfsigma\cdot{\boldsymbol{B}}T}^{(1,0)} = 1 + {\cal O}(\als^2)$ is a matching 
coefficient of the EFT, $c_F = 1 + \dots$ is the matching coefficient of the chromomagnetic 
interaction in the heavy quark effective theory, which is known up to three loops \cite{Grozin:2007fh}, 
$\bfsigma^{(i)}$ is a Pauli matrix acting on the heavy quark labeled $i$ and $T^a_{\bar{3}}$
are the Gell-Mann matrices in the $\bar{3}$ representation.
From Eq. \eqref{QQhfs}, it follows that the hyperfine splitting between the 
$S$-wave ground state of a doubly heavy baryon of spin $1/2$ ($\Xi_{QQ}$) and
the corresponding state of spin $3/2$ ($\Xi_{QQ}^{*}$) may be related to 
the hyperfine splitting between the $S$-wave ground state of a heavy-light meson of spin 
0 ($P_{Q'}$) and the corresponding state of spin 1 ($P_{Q'}^*$):
\begin{eqnarray}
M_{\Xi^*_{QQ}}-M_{\Xi_{QQ}} = \frac{3\, m_{Q'}}{4\, m_Q}\frac{c_F^{(Q)}}{c_F^{(Q')}} 
\left(M_{P^*_{Q'}}-M_{P_{Q'}}\right)  
\left[1+{\cal O}\left(\als^2, \frac{\lQ}{m_Q} , \frac{\lQ}{m_{Q'}} \right)\right],
\end{eqnarray}
where we have kept different the mass, $m_Q$, of the heavy quarks in the baryons 
from the one, $m_{Q'}$, in the mesons.
The obtained figures compare well with existing lattice determinations (see discussion and 
references in ~\cite{Brambilla:2005yk}, a more recent unquenched lattice determination 
of $M_{\Xi^*_{bb}}-M_{\Xi_{bb}}$, which is consistent with previous quenched  determinations, 
may be found in~\cite{Lewis:2008fu}). 

Other terms in the effective Lagrangian have been derived in \cite{Brambilla:2005yk}.

\section{EFT for $QQQ$}
\label{sec_QQQ}
The EFT Lagrangian that describes $QQQ$ baryons below the momentum transfer scale $p$, assumed 
to be larger than $\lQ$, has the general form of Eq. \eqref{lagrangian}, with 
$\delta\mathcal{L}^{(i,j)}$ made out of six quark fields~\cite{Brambilla:2005yk}. 

The term $\delta\mathcal{L}^{(0,0)}$ is 
\begin{equation}
\delta\mathcal{L}^{(0,0)} = \int d^3 r_1\,d^3r_2 \;
\: S^\dagger \left[ i\partial_0 - V^{(0)}_S \right] S 
+ O^{\dagger} \left[ iD_0 - V^{(0)}_{O} \right] O 
+ \Delta^\dagger \left[ iD_0 - V^{(0)}_{\Delta} \right] \Delta ,
\end{equation}
where $S$ is the singlet field, $\displaystyle O=\binom{O^{\rm A}}{O^{\rm S}}$, with 
$O^{\rm A}=(O^{\rm A\,1},O^{\rm A\,2},\dots, O^{\rm A\,8})$ and  
$O^{\rm S}=(O^{\rm S\,1},O^{\rm S\,2},\dots, O^{\rm S\,8})$, 
are the fields that parameterize the two possible octet configurations of three quarks, 
$\Delta = (\Delta^1, \Delta^2,$ $\dots, \Delta^{10})$ are the ten independent $QQQ$ decuplet 
fields and the gauge fields in the covariant derivatives acting on the 
octets and decuplet fields are understood in the $8$ and $10$ representations respectively.
The matching coefficients $V_S^{(0)}$, $V_O^{(0)}$ and $V_\Delta^{(0)}$ can be identified 
with the singlet, octet and decuplet static potentials respectively; 
note that the octet potential $V_O^{(0)}$ is a $2\times 2$ matrix.
The $QQQ$ singlet static potential $V_S^{(0)}$ has been calculated up to 
NNLO in~\cite{Brambilla:2009cd}; it reads
\begin{eqnarray}
V_S^{(0)} \!\! &=& \!\! -\frac{2}{3}
\sum_{q=1}^{3}\frac{\als(1/\vert\br_q\vert)}{\vert\br_q\vert}
\left\{1+\frac{\als(1/\vert\br_q\vert)}{4\pi}\left[
\frac{31}{3}+22 \gamma_E -\left(\frac{10}{9}+\frac{4}{3}\gamma_E\right)n_f\right]
\right.
\nonumber\\
&& 
\hspace{5mm}
\left.
+\left(\frac{\als}{4\pi}\right)^2
\left[66 \zeta(3)+484 \gamma_E^2+ \frac{1976}{3}\gamma_E + \frac{3}{4}\pi^4 + \frac{121}{3}\pi^2 + \frac{4343}{18}
\right.\right.
\nonumber\\
&& \hspace{25mm}
-\left(\frac{52}{3} \zeta(3) + \frac{176}{3}\gamma_E^2 +\frac{916}{9}\gamma_E 
+ \frac{44}{9}\pi^2 + \frac{1229}{27}\right)n_f 
\nonumber\\
&& \hspace{25mm}
\left.\left.
+\left(\frac{16}{9}\gamma_E^2 + \frac{80}{27}\gamma_E + \frac{4}{27}\pi^2
+\frac{100}{81}\right)n_f^2
\right]
\right\}
\nonumber\\
&&\!\!
+ V^\mathrm{3body}_S(\br_1,\br_2,\br_3),
\end{eqnarray}
where $\br_1$, $\br_2$ and $\br_3$ are the distances between the heavy quarks; only two of them 
are independent: if we call $\bx_1$, $\bx_2$ and $\bx_3$ the coordinates of the three quarks, our choice is 
$\br_1=\bx_1-\bx_2$, $\br_2=\bx_1-\bx_3$ and $\br_3=\bx_2-\bx_3$, which implies that $\br_3 = \br_2-\br_1$.
$V^\mathrm{3body}_S$ is the three-body part of the perturbative potential, 
defined as the part of the potential that vanishes when putting one of the quarks 
at infinite distance from the other two.

\begin{figure}
\makebox[0truecm]{\phantom b}
\put(8,0){\epsfxsize=5.5truecm \epsfbox{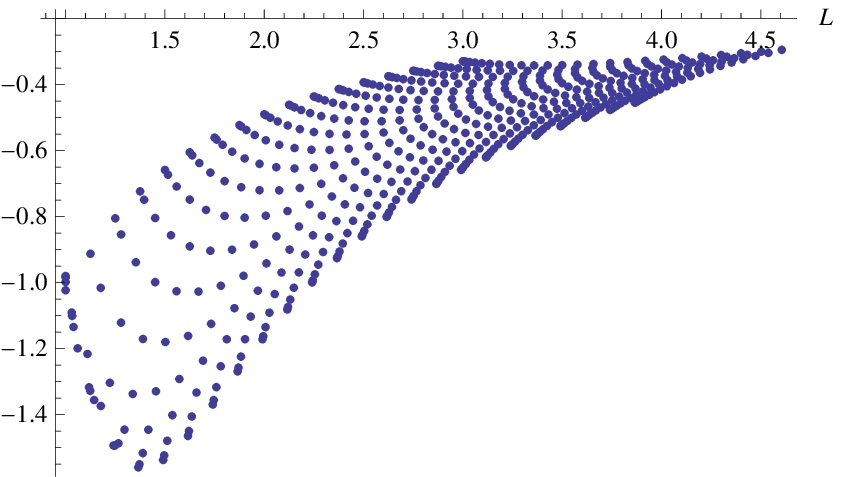}}
\put(188,0){\epsfxsize=5.5truecm \epsfbox{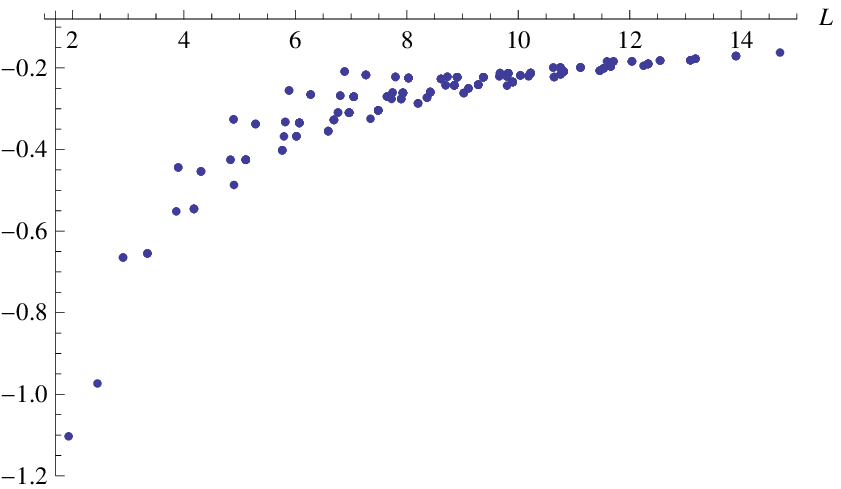}}
\put(0,90){\bf (a)}
\put(180,90){\bf (b)}
\caption{
The normalized three-body potential, $2 V^\mathrm{3body}_S/\als^3$, plotted as function of $L$ 
in arbitrary units for the two geometries described in the text.}
\label{fig:threebody}
\end{figure}

The three-body part is a specific feature of the two-loop potential; it reads
\begin{equation}
V^\mathrm{3body}_S = 
-\als \left(\frac{\als}{4\pi}\right)^2\left[v(\br_2,\br_3)+v(\br_1,-\br_3)+v(-\br_2,-\br_1)\right],
\end{equation}
where 
\begin{eqnarray}
v(\bfrho,\bflambda) &=& 16\pi\hat{\bfrho}\cdot \hat{\bflambda}
\int_0^1dx\int_0^1dy\,\frac{1}{R}
\left[\left(1-\frac{M^2}{R^2}\right)\arctan\frac{R}{M}+\frac{M}{R}\right]
\nonumber\\
&& + 16\pi \hat{\bfrho}^i\hat{\bflambda}^j 
\int_0^1dx\int_0^1dy \, \frac{\hat{\brg}^i\hat{\brg}^j}{R}
\left[\left(1+3\frac{M^2}{R^2}\right)\arctan\frac{R}{M}-3\frac{M}{R}\right],
\end{eqnarray}
with 
$\brg=x\bfrho-y\bflambda$, $R =|\brg|$ and $M=\vert\bfrho\vert\sqrt{x(1-x)}+\vert\bflambda\vert\sqrt{y(1-y)}$.
This term has a different dependence on the positions of the three quarks 
with respect to the Coulomb potential.
It is finite for all configurations: it vanishes when one of the quarks is pulled at infinite distance 
(note that in this case $V_S^{(0)}$ becomes $V_T^{(0)}$) and still remains finite when two quarks are put 
in the same position (note that in this case $V_S^{(0)}$ becomes the $Q\bar{Q}$ singlet static potential).
Hence, its dependence on the geometry is much smoother than the dependence on the geometry 
of the Coulomb potential, although it still clearly depends on it. 
These features can be seen by plotting $V^\mathrm{3body}_S$ as a function of 
$L$ for two different set of configurations; $L$ is the sum of the distances of the three quarks 
from the so-called Fermat (or Torricelli) point, which has minimum distance from the quarks. 
Figure~\ref{fig:threebody}(a) shows the three-body potential when  
we place the three quarks in a plane $(x,y)$,  
fix the position of the first quark in  $(0,0)$, the position of the second one  
in $(1,0)$ and move the third one in $(0.5 + 0.125\,n_x, 0.125\,n_y)$  
for $n_x\in \{0,1, \dots,  20\}$ and  $n_y\in \{0,1, \dots,  24\}$.
The plot shows a clear dependence on the geometry at fixed $L$ (i.e. different configurations 
with same $L$ give different potentials), although weaker than in the Coulombic case.
Figure~\ref{fig:threebody}(b) shows the three-body potential in a geometry used in ~\cite{Takahashi:2000te}, 
which consists in placing the three quarks 
along the axes of a three-dimensional lattice: $(n_x,0,0)$,  $(0,n_y,0)$ and  $(0,0,n_z)$, 
for $n_x \in \{0,1, \dots, 6\}$ and $n_y,n_z\in \{1, \dots, 6\}$.
The plot shows a weaker dependence on the geometry than in the previous case. 
As the comparison with Fig.~\ref{fig:threebody}(a) indicates, 
the weaker dependence is an artifact of the chosen configurations  
rather than a physical effect. Clearly, the precise identification 
of the transition region from a two-body dominated potential 
to a three-body dominated one would require to perform lattice calculations in 
geometries different from the one used in Fig.~\ref{fig:threebody}(b).
The use of different geometries could also be important 
to assess the nature of the long-range three-body potential.

The one-gluon exchange mixes the octet fields, 
so that the octet potential $V_O^{(0)}$ is a non-diagonal $2\times 2$ matrix
already at leading order. Choosing $O^{\rm S}$ and $O^{\rm A}$ to be respectively 
symmetric and antisymmetric for exchanges of the quarks located in $\bx_1$ and $\bx_2$
(a different choice of the octet fields would correspond to a field redefinition leading to 
a different octet potential), we obtain
\begin{equation}
V_{O}^{(0)}=\als\left[\frac{1}{|\boldsymbol{r}_1|} 
\left(\begin{array}{cc}-\frac{2}{3}&0\\
0&\frac{1}{3}\end{array}\right)+\frac{1}{|\boldsymbol{r}_2|}
\left(\begin{array}{cc}\frac{1}{12}&-\frac{\sqrt{3}}{4}\\
-\frac{\sqrt{3}}{4}&-\frac{5}{12}\end{array}\right) 
+\frac{1}{|\boldsymbol{r}_3|}
\left(\begin{array}{cc}\frac{1}{12}&\frac{\sqrt{3}}{4}\\ 
\frac{\sqrt{3}}{4}&-\frac{5}{12}\end{array}\right)\right], 
\end{equation}
while at leading order the decuplet potential is 
\begin{equation}
V_{\Delta}^{(0)}= \frac{\als}{3} \,
\left( \frac{1}{|\boldsymbol{r}_1|} 
+ \frac{1}{|\boldsymbol{r}_2|}
+ \frac{1}{|\boldsymbol{r}_3|} 
\right). 
\end{equation}
We mention that there exist lattice data that show very clearly the singlet, octet and decuplet 
$QQQ$ static potentials, although in an equilateral geometry, 
where the two octets are degenerate~\cite{Hubner:2007qh}. 

Other terms in the effective Lagrangian have been derived in \cite{Brambilla:2005yk}.

\section{Conclusions}
\label{sec_conclu}
Systems made of two or three heavy quarks or antiquarks develop similar hierarchies 
of energy scales and may be treated in similar EFT frameworks.

In the case of $Q\bar{Q}$ mesons, the static potential has been determined up to 
next-to-next-to-next-to-leading order in perturbation theory and to a high 
accuracy on the lattice. The Coulomb behaviour starts getting substantial modifications  
at distances around 0.2 fm turning over a linearly raising potential 
at larger distances. Therefore the transition region 
can be studied to a large extent with perturbative methods.
Terms proportional to powers of $1/m$ and $r$ 
have been also calculated and included systematically in physical observables. 

In the case of $QQq$ baryons, the static potential has been determined up to 
NNLO in perturbation theory and recently also on the lattice. 
Terms proportional to powers of $1/m$ and $r$ in the Lagrangian have been matched 
(mostly) at leading order and used to determine, for instance, the expected 
hyperfine splitting of the ground state of these systems. 
If early experimental evidences will get confirmation in future experimental facilities, 
this will mark the beginning of a future new spectroscopy.

Finally, in the case of $QQQ$ baryons, the static potential has been determined up to 
NNLO in perturbation theory and also on the lattice. The transition region from 
a Coulomb to a linearly raising potential is characterized in this case also 
by the emergence of a three-body potential apparently parameterized by only one length. 
While we have argued that more lattice studies employing different geometries would be 
necessary to precisely identify the transition region, we have also shown that in 
perturbation theory a smooth genuine three-body potential shows up at two loops.

\begin{acknowledgements}
I  acknowledge financial support from the RTN Flavianet MRTN-CT-2006-035482 (EU) 
and from the DFG cluster of excellence ``Origin and structure of the universe'' 
(http://www.universe-cluster.de). 
\end{acknowledgements}

\end{document}